\documentclass[twocolumn,prl,aps,citeautoscript,showpacs,footinbib,amsmath,amssymb,superscriptaddress,longbibliography]{revtex4-2}
\usepackage{array}
\usepackage{graphicx}
\usepackage{dcolumn}
\usepackage{color}
\usepackage{bm}
\usepackage{natbib}
\usepackage{hyperref}
\usepackage{amsmath,amssymb,amsfonts,bbold}
\usepackage[normalem]{ulem}
\usepackage{float}

\usepackage{tikz}

\newcommand{\beq}{\begin{equation}}
\newcommand{\eeq}{\end{equation}}
\newcommand{\beqa}{\begin{eqnarray}}
\newcommand{\eeqa}{\end{eqnarray}}

\usepackage{xcolor}

\usepackage{soul}

\begin{document}

\title{Experimental Realization of the Topologically Nontrivial Phase in Monolayer Si$_2$Te$_2$}

%
%
%

\author{Xiaochun Huang$^\dag$}	
\thanks{These authors contributed equally.}
\email[corresponding author: ]{xiaochun.huang@uni-wuerzburg.de} 
	\affiliation{Physikalisches Institut, Experimentelle Physik II, Universit\"{a}t W\"{u}rzburg, Am Hubland, 97074 W\"{u}rzburg, Germany}
\author{Lingxiao Zhao$^\ast$} 
	\affiliation{Quantum Science Center of Guangdong-Hong Kong-Macao Greater Bay Area (Guangdong), Shenzhen, China} 
\author{Rui Xiong$^\ast$}
\affiliation{Multiscale Computational Materials Facility, Key Laboratory of Eco-materials Advanced Technology, 
		College of Materials Science and Engineering, Fuzhou University, Fuzhou, PR China} 
\author{Wenbin~Li}
\affiliation{The Institute for Solid State Physics, The University of Tokyo; Chiba 277-8581, Japan}
\author{Bao-tian Wang}
\affiliation{Institute of High Energy Physics, Chinese Academy of Sciences, Beijing 100049, China}
\affiliation{Spallation Neutron Source Science Center, Dongguan 523803, China}
\author{Baisheng Sa}
\email[corresponding author: ]{bssa@fzu.edu.cn}
\affiliation{Multiscale Computational Materials Facility, Key Laboratory of Eco-materials Advanced Technology, 
		College of Materials Science and Engineering, Fuzhou University, Fuzhou, PR China} 
\author{Matthias Bode}
\affiliation{Physikalisches Institut, Experimentelle Physik II, Universit\"{a}t W\"{u}rzburg, Am Hubland, 97074 W\"{u}rzburg, Germany}	
\affiliation{Wilhelm Conrad R{\"o}ntgen-Center for Complex Material Systems (RCCM), 
		Universit\"{a}t W\"{u}rzburg, Am Hubland, 97074 W\"{u}rzburg, Germany}    

	   

\date{\today}

\begin{abstract}
The free-standing monolayer Si$_2$Te$_2$ (ML-Si$_2$Te$_2$) 
has been theoretically predicted to host a room-temperature quantum spin Hall phase. 
However, its experimental realization remains challenge due to the absence of a three-dimensional counterpart. 
Here, we demonstrate that HfTe$_2$ serves as an ideal substrate 
for the epitaxial growth of ML-Si$_2$Te$_2$, preserving its topological phase. 
Scanning tunneling microscopy and spectroscopy confirm a strain-free ${(1 \times 1)}$ lattice of ML-Si$_2$Te$_2$, 
along with a sizable band gap, which is well captured by first-principles calculations. 
Moreover, distinct edge states, independent of step geometry and exhibiting a broad spatial distribution, 
are observed at ML-Si$_2$Te$_2$ step edges, underscoring its topological nature.
\end{abstract}


\maketitle


Quantum spin Hall (QSH) insulators, also known as two-dimensional (2D) topological insulators (TIs), 
host one-dimensional (1D) helical edge states with Dirac-like linear dispersion 
which are dictated by the bulk band topology \cite{R1,R2,R3,R4}. 
Protected by time-reversal symmetry, these edge states are immune to backscattering, 
offering potential applications in low-energy consumption devices \cite{R3,R4,R5}. 
While 2D TI phases have been experimentally demonstrated in HgTe/CdTe and InAs/GaSb quantum wells \cite{R6,R7}, 
their small bulk band gaps---typically on the order of meV---limit both further experimental studies and practical applications. 
In recent years, extensive efforts have been devoted to identify new 2D TIs with sizable band gaps, 
tapping into the vast potential of emerging 2D materials \cite{R8,R9,R10}. 

A common strategy is to realize 2D TI phases by reducing three-dimensional (3D) materials to the monolayer (ML) limit, 
as demonstrated in WTe$_2$ \cite{R11,R12,R13}, ZrTe$_5$ \cite{R14,R15}, 
and Bi$_4$Br$_4$ monolayers \cite{R16,R17}, though with limited success. 
Alternatively, a variety of 2D TIs with artificial lattice structures have been theoretically proposed, 
enabled by specific lattice symmetries and strong spin-orbit coupling (SOC), 
thereby broadening the range of potential candidates \cite{R18}. 
However, among these, only a few, primarily graphene-like elementary 2D materials, 
such as germanene \cite{R19}, stanene \cite{R20,R21}, and bismuthene \cite{R22,R23,R24}, 
have been successfully synthesized and confirmed as topologically nontrivial. 
Furthermore, substantial substrate interactions are often necessary to stabilize their artificial lattices \cite{R25,R26}.

Monolayer Si$_2$Te$_2$ (ML-Si$_2$Te$_2$) possesses an artificial lattice structure 
and is predicted to exhibit a room-temperature QSH phase \cite{R27,R28}. 
This novel 2D material features a hexagonal ($P\overset{-}{3}$\,m1) symmetry with a unique Te--Si--Si--Te stacking sequence, 
suggesting that only weak van der Waals (vdW) interactions with the substrate may be required \cite{R28}. Density functional theory (DFT) calculations indicate that free-standing ML-Si$_2$Te$_2$ 
has a topologically nontrivial band gap of 220\,meV, while its band topology is highly sensitive to lattice strain \cite{R28,R29}. 
Since ML-Si$_2$Te$_2$ has no 3D counterpart, it cannot be obtained via mechanical exfoliation from a bulk material. 
Recently, we successfully achieved 
the epitaxial growth of ML-Si$_2$Te$_2$ on an Sb$_2$Te$_3$ substrate \cite{R29,R30}. 
Nevertheless, the large lattice mismatch between free-standing ML-Si$_2$Te$_2$ and Sb$_2$Te$_3$ 
induces significant interfacial strain, deriving the system into a trivial semiconducting phase \cite{R29,R30}.

In this work, by combining DFT calculations, molecular beam epitaxy (MBE), 
and scanning tunneling microscopy/spectroscopy (STM/STS), 
we report the first experimental realization of the QSH phase in ML-Si$_2$Te$_2$ grown on a HfTe$_2$ substrate. 
We demonstrate that ML-Si$_2$Te$_2$ interacts with HfTe$_2$ via vdW forces, 
and that its lattice constants match those of free-standing ML-Si$_2$Te$_2$, indicating a strain-free ${(1 \times 1)}$ structure. 
The sizable nontrivial bulk band gap predicted by DFT calculations 
is confirmed by differential tunneling conductivity (d$I$/d$V$) measurements. 
More importantly, distinct topological edge states are observed within the band gap of ML-Si$_2$Te$_2$. 
These findings pave the way for further exploration of the potential room-temperature QSH effect in this novel 2D material.

\begin{figure*}[t]
	\centering
	\includegraphics[width=\textwidth,angle=0,scale=1]{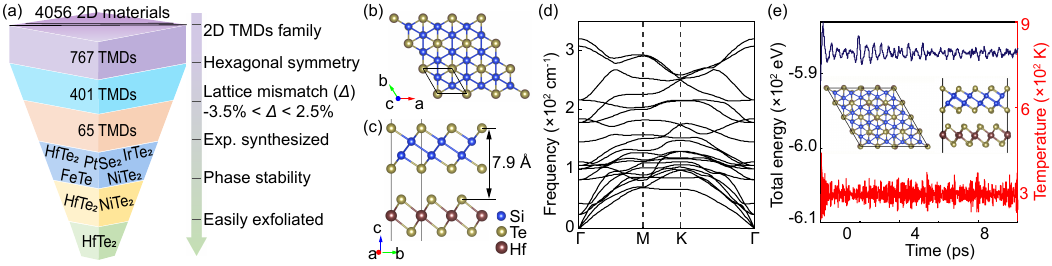}
	\caption{{\bf ML-Si$_2$Te$_2$ on HfTe$_2$.} 
		(a) Screening process of TMDs as substrates for the epitaxial growth of ML-Si$_2$Te$_2$. 
		(b,c) Top and side view, respectively, of the structural model of ML-Si$_2$Te$_2$ on HfTe$_2$, 
		with the unit cell outlined by a diamond and vertical lines. 
		(d) Calculated phonon dispersion curves of the ML-Si$_2$Te$_2$/HfTe$_2$ heterostructure. 
		(e) Main panel: AIMD evolution of total energy and temperature 
		for the ML-Si$_2$Te$_2$/ML-HfTe$_2$ heterostructure with a  ${4 \times 4}$ supercell. 
		Inset: Snapshot of the ML-Si$_2$Te$_2$/HfTe$_2$ heterostructure after $10 \text{ ps}$ of simulation at $300 \text{ K}$.}
	\label{Figure1}
\end{figure*}

To identify a suitable substrate, we screened candidates from the Computational 2D Materials Database (C2DB) \cite{R31,R32}, 
which contains 4056 2D materials, see Fig.\,\ref{Figure1}(a). 
Transition metal chalcogenides (TMDs) were chosen as the initial screening criterion due to their structural diversity. Among the 767 TMDs, 401 exhibit trigonal (or hexagonal) symmetry, matching that of ML-Si$_2$Te$_2$. 
Our previous work shows that maintaining the topological phase of ML-Si$_2$Te$_2$ 
requires an in-plane strain within $-3$\% to 2\% relative to the free-standing case ($a_0 = b_0 = 389$\,pm) \cite{R29}. 
To account for vdW interactions, which can enhance lattice mismatch tolerance, 
we adopted a slightly broader criterion of $-3.5$ to 2.5\%, defined as $(a_0 - a_{\text{TMD}})/a_0$, yielding 76 candidates. 
Filtering for materials which have been successfully synthesized narrows the selection down to five candidates: 
IrTe$_2$, NiTe$_2$, FeTe, PtSe$_2$, and HfTe$_2$. 
IrTe$_2$ transitions to a triclinic phase below $\approx 270$\,K \cite{R33} and hexagonal FeTe is stable only above 983\,K \cite{R34}, 
making both unsuitable for low-temperature STM/STS measurements. 
PtSe$_2$ was excluded to prevent potential Se-Te substitution or alloying during Si$_2$Te$_2$ growth. 
To ensure an atomically flat surface via cleaving in ultrahigh vacuum (UHV), 
we evaluated the exfoliation energies of NiTe$_2$, and HfTe$_2$. 
As shown in Fig.\,S1 of Ref.\,\onlinecite{SupplMat}, 
HfTe$_2$ exhibits an apparently lower exfoliation energy than NiTe$_2$, indicating easier cleavage. 
Based on these considerations, we selected HfTe$_2$ as the substrate. 

The structural stability of ML-Si$_2$Te$_2$ on HfTe$_2$ was investigated via DFT calculations 
employing the Vienna {\em ab initio} simulation package (VASP) \cite{R35,SupplMat}. 
Fig.\,\ref{Figure1}(b) and \ref{Figure1}(c) illustrate the ML-Si$_2$Te$_2$/ML-HfTe$_2$ heterostructure,
which serves as the model configuration for our calculations. 
After full structural relaxation, the in-plane lattice constants of ML-Si$_2$Te$_2$ remain at 
$a_{\text{DFT}} = b_{\text{DFT}} = 389$\,pm, identical to those of free-standing case, with an interlayer spacing of $790$\,pm. 
The calculated binding energy of $-0.296$\,eV/unit cell indicates a typical vdW interaction \cite{SupplMat}. 
Fig.\,\ref{Figure1}(d) presents the phonon dispersion curves, 
where the absence of imaginary frequencies confirms the dynamic stability of the heterojunction. 
Further {\em ab initio} molecular dynamics (AIMD) simulations demonstrated its thermal stability \cite{R36}. 
As shown in Fig.\,\ref{Figure1}(e), the total energy fluctuates within a small range, 
and the lattice remains stable during annealing at $300$\,K for $10$\,ps. 

\begin{figure}[b]
	\centering
	\includegraphics[width=\columnwidth,angle=0,scale=1]{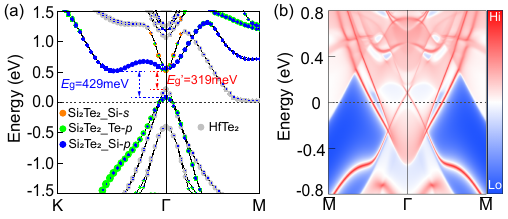}
	\caption{{\bf Calculated band structures of  ML-Si$_2$Te$_2$ on HfTe$_2$ with SOC}. 
		(a) Projected band structure, where the symbol size represents the orbital contribution. 
		(b) Spectral density of the edge states. Gray dashed lines in (a) and (b) indicate the Fermi levels.}
	\label{Figure2}
\end{figure}

Next, electronic structure calculations were performed using the Heyd-Scuseria-Ernzerhof (HSE06) hybrid functional 
to investigate the band topology of ML-Si$_2$Te$_2$ on HfTe$_2$ \cite{R37}. 
To assess the substrate effects, comparative calculations were conducted for the free-standing case \cite{SupplMat}. 
The decomposed band structures [Fig.\,\ref{Figure2}(a) and Fig.\,S2] reveal that the projected band dispersion 
of ML-Si$_2$Te$_2$ in the heterojunction closely resembles that of the free-standing case 
and preserves the SOC-induced band inversion at the $\varGamma$ point, 
indicating that the topologically nontrivial phase remains intact. 
Notably, substrate interactions shift the conduction band minimum (CBM) of ML-Si$_2$Te$_2$ upward, 
enhancing its nontrivial band gap to 429\,meV. 
Given that the valance band maximum (VBM) of ML-HfTe$_2$ lies 110\,meV above that of ML-Si$_2$Te$_2$, 
the experimentally observed band gap at the $\varGamma$ point is expected to be ${\approx}319$\,meV. 
Besides, the Fermi level ($E_{\text{F}}$) is shifted downwards below the VBM of ML-Si$_2$Te$_2$ 
because of the higher work function of ML-HfTe$_2$ relative to ML-Si$_2$Te$_2$ (see Ref.\,\onlinecite{SupplMat} for details). 
Fig.\,\ref{Figure2}(b) presents the calculated local density of states (LDOS) for the edge states, revealing a linear dispersion at $\varGamma$. 
Comparison with the free-standing case (Fig.\,S3) confirms their Dirac-like dispersion. 
The nontrivial topology is further verified by calculating the $Z_2$ invariant 
via the evolution of Wannier charge centers (WCCs) \cite{R38}. 
As shown in Fig.\,S3 in Ref.\,\onlinecite{SupplMat}, the WCCs curves for both the heterostructure 
and the free-standing ML-Si$_2$Te$_2$ cross an arbitrary horizontal reference line 
an odd number of times, confirming a $Z_2 = 1$ in both cases. 
Our DFT results indicate that strain-free ML-Si$_2$Te$_2$ 
with a QSH phase can be realized on a HfTe$_2$ substrate.

\begin{figure}[t]
	\centering
	\includegraphics[width=\columnwidth,angle=0,scale=1]{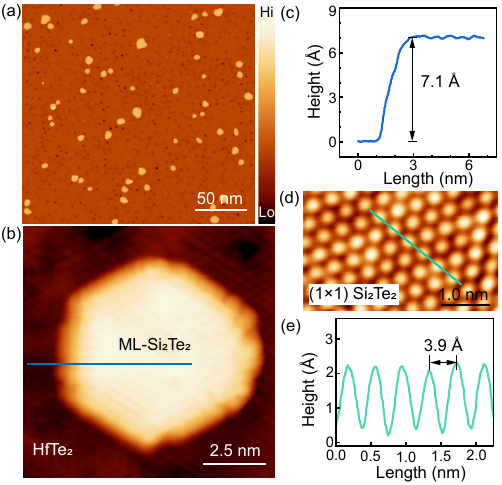}
	\caption{{\bf Morphology of ML-Si$_2$Te$_2$ grown on HfTe$_2$.} 
		(a) Large-scale STM image ($V = 1.0$\,V, $I = 10$\,pA) of ML-Si$_2$Te$_2$ grown on a HfTe$_2$ substrate. 
		(b) High-resolution STM image ($V = 0.5$\,V, $I = 200$\,pA) showing the surface lattice structure 
		of both ML-Si$_2$Te$_2$ and the HfTe$_2$ substrate simultaneously. 
		(c) Height profile taken along the blue line in (b). 
		(d) Atomic-resolution STM image ($V = 0.5$\,V, $I = 100$\,pA) of ML-Si$_2$Te$_2$. 
		(e) Line profile, taken along the green line in (d).}
	\label{Figure3}
\end{figure}

High-quality HfTe$_2$ single crystals, with lateral dimensions of several millimeters, 
were synthesized via chemical vapor transport, using I$_2$ as the transport agent, 
in a two-zone furnace (see Fig.\,S4 and Ref.\,\onlinecite{SupplMat} for details). 
Epitaxial growth and electronic characterization of ML-Si$_2$Te$_2$ on HfTe$_2$ were performed 
in a two-chamber UHV-MBE system with a base pressure $p < 3 \times 10^{-11} \text{ Torr}$ and an integrated home-built low-temperature STM. 
{\em In-situ} STM/STS measurements were conducted 
at 5.8 K with W-tips (see Ref.\,\onlinecite{SupplMat} for details). 
Atomically smooth HfTe$_2$ surfaces were obtained by cleaving the crystals 
under UHV conditions (see Fig.\,S5 in Ref.\,\onlinecite{SupplMat}). 
Monolayer Si$_2$Te$_2$ was fabricated by co-evaporating high-purity Si ($99.999\%$) and Te ($99.9999\%$) 
onto the HfTe$_2$ substrate at $200^{\circ}$C, followed by post-annealing at elevated temperature. 
Annealing studies detailed in Fig.\,S6 of Ref.\,\onlinecite{SupplMat} show that 
high-quality ML-Si$_2$Te$_2$ is achieved at $385^{\circ}$C, although a 
significant reduction in coverage is observed above $320^{\circ}$C---well below the crystallization threshold of ML-Si$_2$Te$_2$. 
To counteract the decomposition of SiTe$_x$ and maintain sufficient precursor material on the substrate, 
we continuously supplied both Si and Te fluxes exceeding $20$ times the nominal co-evaporation rate during annealing, thereby enabling the successful growth of ML-Si$_2$Te$_2$ (see Ref.\,\onlinecite{SupplMat} for details).

Figure\,\ref{Figure3}(a) shows a large-scale STM image of the as-grown sample, providing an overview of its morphology. 
A close-up STM image of an island [Fig.\,\ref{Figure3}(b)] reveals the hexagonal surface structure 
of both Si$_2$Te$_2$ and the underlying HfTe$_2$ substrate simultaneously. 
The step height of $710$\,pm, extracted from the line profile in Fig.\,\ref{Figure3}(c), 
is in reasonable agreement with the calculated interlayer spacing \cite{R29,R39}. 
Atomic-resolution STM imaging [Fig.\,\ref{Figure3}(d)] yields in-plane lattice constants 
of ML-Si$_2$Te$_2$ ($a_{\text{exp}} = b_{\text{exp}} = 390$\,pm), in excellent agreement 
with theoretical values ($a_{\text{DFT}} = b_{\text{DFT}} = 389$\,pm). Systematic measurements across multiple samples consistently reveal a well-defined epitaxial relationship, 
where ML-Si$_2$Te$_2$ islands align with the HfTe$_2$ substrate in a hollow-site configuration, 
consistent with DFT calculations (see Fig.\,S7 and Fig.\,S8 in Ref.\,\onlinecite{SupplMat}). 
These results show clear evidence for the experimental realization of strain-free ${(1 \times 1)}$ ML-Si$_2$Te$_2$.

\begin{figure}[h]
	\centering
	\includegraphics[width=\columnwidth,angle=0,scale=1]{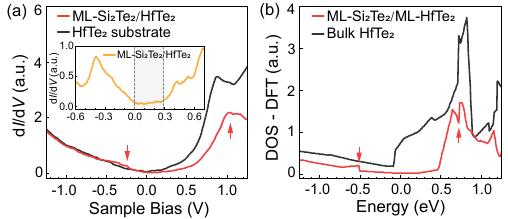}
	\caption{{\bf Local density of states of ML-Si$_2$Te$_2$ on HfTe$_2$.} 
		(a) Main panel: Local d$I$/d$V$ spectra ($V_{\text{stab}} = -1.5$\,V, $I_{\text{stab}} = 200$\,pA, 
		and $V_{\text{mod}} = 10$\,mV) measured on ML-Si$_2$Te$_2$ and the HfTe$_2$ substrate. 
		Inset: High-resolution d$I$/d$V$ spectra ($V_{\text{stab}} = -1.0$\,V, $I_{\text{stab}} = 100$\,pA, 
		and $V_{\text{mod}} = 10$\,mV) acquired on ML-Si$_2$Te$_2$. 
		(b) DFT-calculated projected DOS for ML-Si$_2$Te$_2$ in ML-Si$_2$Te$_2$/ML-HfTe$_2$ and total DOS of bulk-HfTe$_2$.}
	\label{Figure4}
\end{figure}

We investigated the electronic structure of strain-free ML-Si$_2$Te$_2$ by combining STS with DFT calculations. 
Figure\,\ref{Figure4}(a) presents representative d$I$/d$V$ spectra acquired on ML-Si$_2$Te$_2$ and on the HfTe$_2$ substrate. 
The ML-Si$_2$Te$_2$ spectrum is characterized by two prominent peaks separated by ${\sim}1.25 \text{ eV}$ (marked by red arrows). 
As shown in Fig.\,\ref{Figure4}(b), DFT calculations reproduce this feature, albeit with an overall energy shift 
of ${\sim}0.25 \text{ eV}$. This shift is attributed to substrate-induced doping effect that are not fully captured in DFT model \cite{R22,R29}. Taken together, the good agreement between the experimental and theoretical results 
confirms the formation of a well-defined band structure in the synthesized ML-Si$_2$Te$_2$ islands.

Since a bulk band gap is essential for the QSH phase \cite{R13,R22}, the experimental confirmation of the band gap 
in ML-Si$_2$Te$_2$ grown on HfTe$_2$---predicted theoretically in Fig.\,\ref{Figure2}(a)---is of particular importance. 
In Fig.\,\ref{Figure4}(a), the red curve exhibits a smooth, low-intensity region spanning from $-100$ to $300$\,meV, 
indicating the presence of a sizable band gap in ML-Si$_2$Te$_2$. 
However, the exact energy positions of the conduction and valence band edges 
appear smeared due to the metallic HfTe$_2$ substrate. 
To better resolve the band edges, we acquired high-resolution d$I$/d$V$ spectra on ML-Si$_2$Te$_2$ 
[see inset in Fig.\,\ref{Figure4}(a), Fig.\,S9(c)], 
from which the VBM ($\approx -10$\,meV) and CBM ($\approx 290$\,meV) can be readily identified (gray dashed lines), defining a band gap of $\approx 300$\,meV. This value is consistent with the DFT-calculated gap of 319 meV [Fig.\,\ref{Figure2}(a)] and corroborated by systematic measurements on 20 ML-Si$_2$Te$_2$ islands (See Ref.\,\onlinecite{SupplMat}).

\begin{figure}[h]
	\centering
	\includegraphics[width=\columnwidth,angle=0,scale=1]{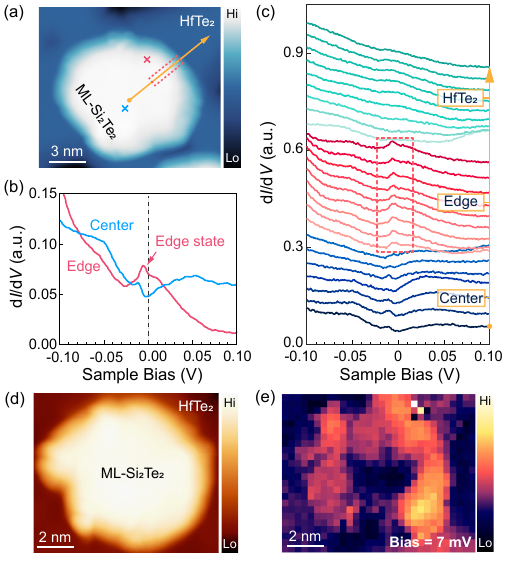}
	\caption{{\bf Identification of the edge states.} 
		(a) STM image ($V = -1.0$\,V, $I = 10$\,pA) of a typical ML-Si$_2$Te$_2$ island. 
		(b) d$I$/d$V$ spectra ($V_{\text{stab}} = -0.1$\,V, $I_{\text{stab}} = 100$\,pA, and $V_{\text{mod}} = 2.0$\,mV) 
		measured at the edge and center of the ML-Si$_2$Te$_2$ island at the positions marked in (a). 
		(c) Spatially dependent d$I$/d$V$ spectra recorded along the origin line in (a) with 0.36~nm spacing. 
		Spectra are vertically offset for clarity. Measurement positions within the red dashed rectangle are indicated in (a).
		(d) STM image ($V = 1.0$\,V, $I = 10$\,pA) of a ML-Si$_2$Te$_2$ island.
		(e) d$I$/d$V$ map ($V_{\text{stab}} = 0.2$\,V, $I_{\text{stab}} = 100$\,pA, and $V_{\text{mod}} = 2.0$\,mV) taken at +7 mV, corresponding to the area shown in (d).}
	\label{Figure5}
\end{figure}

A hallmark of QSH systems is the presence of topological edge states within the bulk band gap, 
which manifest as enhanced d$I$/d$V$ intensity at step edges \cite{R13,R22,R40,R41,R42,R43}. 
Fig.\,\ref{Figure5}(a) and (b) show representative d$I$/d$V$ spectra 
acquired at the step edge (red curve) and the center (blue curve) of a ML-Si$_2$Te$_2$ island. 
The center spectrum exhibits a V-shaped dip between $\pm 50$\,mV, whereas the edge spectrum 
displays a pronounced peak at $E_{\rm F}$ within the same energy range, providing direct evidence of in-gap edge states. 
These features are reproducibly observed across multiple ML-Si$_2$Te$_2$ islands 
and step geometries, see Fig.\,S11, indicating their robust nature.
As discussed in Ref.\,\onlinecite{SupplMat}, 1D topological edge states in ML-Si$_2$Te$_2$ are expected 
to manifest as a d$I$/d$V$ peak near the edge, consistent with the sharp feature observed in our measurements. 

To further investigate the physical origin of these edge states, we acquired a series of d$I$/d$V$ spectra 
along a line spanning from the island center to the HfTe$_2$ substrate, crossing the step edge [Fig.\,\ref{Figure5}(c)].
Spectra obtained at the island center exhibit a minimum in intensity between $\pm 10$\,mV. 
Within this energy range, a small peak is discernible which we attribute to substrate-induced states (see Ref.\,\onlinecite{SupplMat}).
As the measurement position approaches the step edge, this minimum remains essentially unchanged (blue curves), 
until it is overtaken by a pronounced peak that emerges within approximately 2.0 to 3.0\,nm of the edge (red curves) 
and vanishes on the substrate side (green curves). 
The extended spatial profile of this peak supports its topological origin and rules out trivial mechanisms 
such as lattice defects or dangling bonds, which typically produce spatially localized signals confined to just one or two atomic rows \cite{R22}.

Additional evidence is provided by 2D d$I$/d$V$ mapping. 
Figures\,\ref{Figure5}(d) and (e) present a map acquired at the energy of the edge states 
for a representative ML-Si$_2$Te$_2$ island (see Ref.\,\onlinecite{SupplMat} for details).
A ring of enhanced d$I$/d$V$ intensity can be recognized along the island boundary, 
with only minor signal degradation at a small segment of the lower edge, 
likely due to residual tip-related effects (see Ref.\,\onlinecite{SupplMat} for details).
This spatial continuity underscores the robustness of the edge states 
against geometric disorder and further rules out trivial edge models. 
From these spatially resolved data, we estimate the edge-state width to be ${\approx}2$\,nm, 
in good agreement with the DFT-predicted value of ${\approx}1.35$\,nm for ML-Si$_2$Te$_2$ nanoribbons (see Ref.\,\onlinecite{SupplMat}).
The observed spatial extent and robustness of the edge states provide strong evidence for their topological origin.

\smallskip
In summary, we report the first experimental realization of the QSH phase in ML-Si$_2$Te$_2$. 
By combining high-throughput computational screening with band structure calculations, 
we identified HfTe$_2$ as an ideal substrate for the epitaxial growth of ML-Si$_2$Te$_2$, enabling the emergence of topological edge states. 
Subsequent MBE successfully yielded strain-free ${(1 \times 1)}$ ML-Si$_2$Te$_2$ 
on HfTe$_2$ substrates with vdW-type interfacial interactions. 
Using STM/STS, we confirmed a sizable bulk band gap, along with distinct topological edge states within the gap. 
Given that ML-Si$_2$Te$_2$ is a novel artificial 2D material, our findings pave the way 
for further investigations into its topological electronic properties and potential applications in spintronic devices. 
Moreover, this work exemplifies the material-by-design approach in the study of 2D materials and heterostructures.

\smallskip
\begin{acknowledgments}
The authors acknowledge J.\,Qi, P.\,H{\"a}rtl, and J.\,Gou for fruitful discussions. 
X.H.~would like to thank for the financial support from the DFG 
through the Hallwachs-R{\"o}ntgen Postdoc Program of ct.qmat (EXC 2147, Project No. 390858490). 
L.Z.~would like to thank for the financial support from the Guangdong Provincial Quantum Science Strategic Initiative (GDZX2401011). 
\end{acknowledgments}

\end{document}